# AI-assisted Human-in-the-Loop Web Platform for Structural Characterization in Hard drive design

Utkarsh Pratiush[1], Huaixun Huyan[2], Maryam Zahiri Azar[2], Esmeralda Yitamben[2], Allen Bourez[2], Sergei V Kalinin[1], Vasfi Burak Ozdol[2]

[1]Department of Materials Science and Engineering, University of Tennessee, Knoxville, TN 37996, USA

[2]Western Digital corporation, 5601 Great Oaks Parkway, San Jose, CA 95119, USA

*This work was done during WD internship program

## Abstract

Scanning transmission electron microscopy (STEM) has become a cornerstone instrument for semiconductor materials metrology, enabling nanoscale analysis of complex multilayer structures that define device performance. Developing effective metrology workflows for such systems requires balancing automation with flexibility, rigid pipelines are brittle to sample variability, while purely manual approaches are slow and subjective. Here, we present a tunable human-AI-assisted workflow framework that enables modular and adaptive analysis of STEM images for device characterization. As an illustrative example, we demonstrate a workflow for automated layer thickness and interface roughness quantification in multilayer thin films. The system integrates gradient-based peak detection with interactive correction modules, allowing human input at the design stage while maintaining fully automated execution across samples. Implemented as a web-based interface, it processes TEM/EMD files directly, applies noise-reduction and interface-tracking algorithms, and outputs statistical roughness and thickness metrics with nanometer precision. This architecture exemplifies a general approach toward adaptive, reusable metrology workflows—bridging human insight and machine precision for scalable, standardized analysis in semiconductor manufacturing. The code is made available at https://github.com/utkarshp1161/thickness-mapping-webapp

## I. Introduction

Semiconductor device characterization underpins process control and yields learning across the fab, and many metrology pipelines are now sufficiently mature that they are executed repeatedly with only modest adjustments to acquisition or recipe parameters. Representative examples include: (i) Automated patterned/unpatterned wafer defect inspection (optical and e-beam) used for excursion monitoring and yield ramp, and are increasingly coupled to AI classification [1–3], (ii) Critical-dimension (CD) scanning electron microscopy (CD-SEM) for line/space and edge-roughness measurements in lithography stacks[4], (iii) Cross-sectional (S)TEM for nanoscale analysis of multilayer films and device cross-sections, often combined with analytical signals[5], compositional mapping via EDS/EELS for chemistry and bonding state[6], (iv) Depth-profiling by SIMS for dopant and heterostructure profiling[7,8], and, (v) AFM-based topography/roughness and sidewall-roughness metrology for advanced 3D features[9]. Despite this maturity, adapting recipes to new materials and device architectures still demands nontrivial retuning, motivating workflow designs that retain repeatability while allowing targeted, on-the-fly adjustments. Within this landscape, Transmission electron microscopy (TEM)[10] provides unparalleled spatial resolution for analyzing multilayer structures at the nanoscale. However, the quantitative analysis of TEM images traditionally relies on manual measurement techniques that are inherently subjective, time-consuming, and inconsistent between operators. This limitation becomes particularly problematic when analyzing complex multilayer systems with multiple interfaces or when processing large datasets for statistical analysis[11–13].

Among these metrological tasks, important is metrology of multilayer structures. Multilayer structures are fundamental components in modern technology, spanning applications from semiconductor devices and optical coatings to advanced composite materials. The precise characterization of layer thickness and interface roughness is critical for understanding material properties, optimizing manufacturing processes, and ensuring quality control in industrial production[14–16].

In the context of hard-disk-drive (HDD) technology and heat-assisted magnetic recording (HAMR) media in particular, multilayer metrology plays an especially critical role. HAMR disks rely on a sophisticated stack composed of seed and interlayers, thermally engineered underlayers, granular FePt media, segregants, heat-sink layers, overcoats, and in some cases plasmonic near-field transducer (NFT) adjacent layers. The thickness uniformity, crystallographic texture, and interface roughness across these layers directly influence key performance parameters including $L1_0$ ordering, grain size distribution, magnetic anisotropy, thermal gradients, and write efficiency. Even nanometer-scale deviations can degrade coercivity, increase transition noise, or reduce NFT coupling efficiency, ultimately limiting achievable areal density. Equally important is maintaining minimal disk surface roughness, as excessive topography can destabilize the head–disk spacing, increase flying-height modulation, and elevate touchdown risk effects that are even more critical in HAMR, where the near-field transducer operates within a few nanometers of the disk surface. Therefore, precise metrology of HAMR multilayer structures is essential not only for understanding the physical behavior of the media stack but also for guiding manufacturing optimization and ensuring device reliability at the extremely small length scales required for next-generation storage technologies.

Interface roughness measurement presents additional challenges. While atomic force microscopy (AFM) is commonly used for surface roughness characterization, it requires layer-by-layer deposition and measurement for buried interfaces, making it impractical for completed multilayer structures. TEM-based roughness analysis offers the advantage of measuring all interfaces simultaneously in cross-sectional samples, providing comprehensive characterization without destructive sample preparation[15,16]. X-ray reflectivity (XRR) is often used to estimate interface roughness and layer thickness[17], though it lacks the nanoscale spatial resolution of TEM.

Recent advances in artificial intelligence and automated image analysis[18–22] have opened new possibilities for standardized, objective measurement of microscopy images. However, existing solutions often lack the flexibility to handle the diverse range of multilayer structures encountered in practice or require specialized software installations that limit accessibility. For instance, while GUI-based toolkits like CrysTBox[23] deliver powerful automated analysis and visualization, they don't fully address web-based accessibility or thickness/roughness specific workflows.

**Challenge:** Manual EM analysis is slow, error-prone (local measurements); fully automated solutions like using popular segmentation model[24–28] lack robustness in edge cases (e.g., detector noise). Writing custom code is a bottleneck for researchers unfamiliar with programming. There is a need for custom UI-based tools to measure morphology.

We address these challenges by developing a human-in-the-loop[29–31] AI-assisted web-based tool that automates thickness and interface roughness analysis of multilayer structures from TEM images. Our approach combines automated peak-detection algorithms with interactive manual correction capabilities, ensuring both efficiency and accuracy. The web-based implementation provides universal accessibility while maintaining sophisticated analysis capabilities.

**II. Methodology**

Semiconductor metrology benefits from workflows that are repeatable for mature tasks yet flexible for new samples/goals. Fully manual operation is adaptable but slow and operator-dependent; fully rigid automation is fast but brittle. We formalize a tunable human–AI workflow in which (i) humans specify task intent and guardrails (parameters, regions of interest, acceptance criteria) and (ii) automation executes deterministically with live feedback and minimal interventions. We implement, deploy, and validate this approach on STEM/TEM cross-section images, focusing on multilayer thickness and interface roughness.

*Though the method is sample agnostic we used it here for HDD device structures, synthesis details not shared due to prosperity reasons*

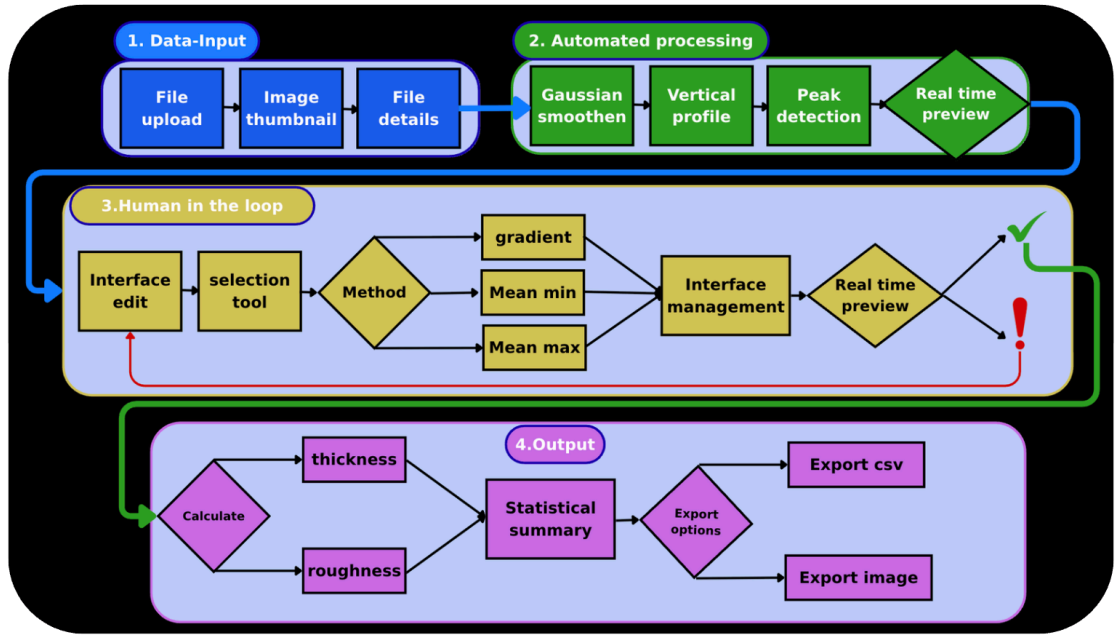

**Figure. 1**: Complete analysis workflow showing the process from EMD file upload through interface detection, thickness calculation, and roughness analysis to final data export. The flowchart illustrates both automated processing steps and user interaction points.

The tool is implemented as a Flask-based web application with a Python backend leveraging scientific computing libraries including HyperSpy for EMD file handling, SciPy for signal processing, and OpenCV for image processing operations. The client-server architecture enables real-time image processing while maintaining responsive user interaction. **Figure 1** illustrates the complete analysis workflow, from EMD file upload through final data export. The system processes files up to 500 MB in size and automatically extracts pixel size calibration from EMD metadata when available. For details on the workflow see Supplementary material. A detailed computational workflow was developed to extract interface and thickness metrics from cross-sectional images. The pipeline includes image normalization, Gaussian smoothing, gradient-based interface detection, and adaptive peak identification to ensure robustness across varying contrast conditions. Quantitative parameters such as interface roughness and layer thickness are computed using geometric tracking and statistical averaging. Complete algorithmic details, parameter definitions, and mathematical formulations are provided in the **Supplementary Information (Sections A–D)**.

- Preprocessing: intensity normalization and Gaussian smoothing **(see SI A).**
- Interface detection: gradient-based profiling with percentile thresholds and distance constraints **(see SI B).**

- Roughness metrics: geometric tracking to report $R_{\text{RMS}}$, $R_a$, and peak-to-valley (see SI C).
- Thickness: adjacent interface spacing converted to physical units via pixel size (see SI D).

## III. Results and Discussion

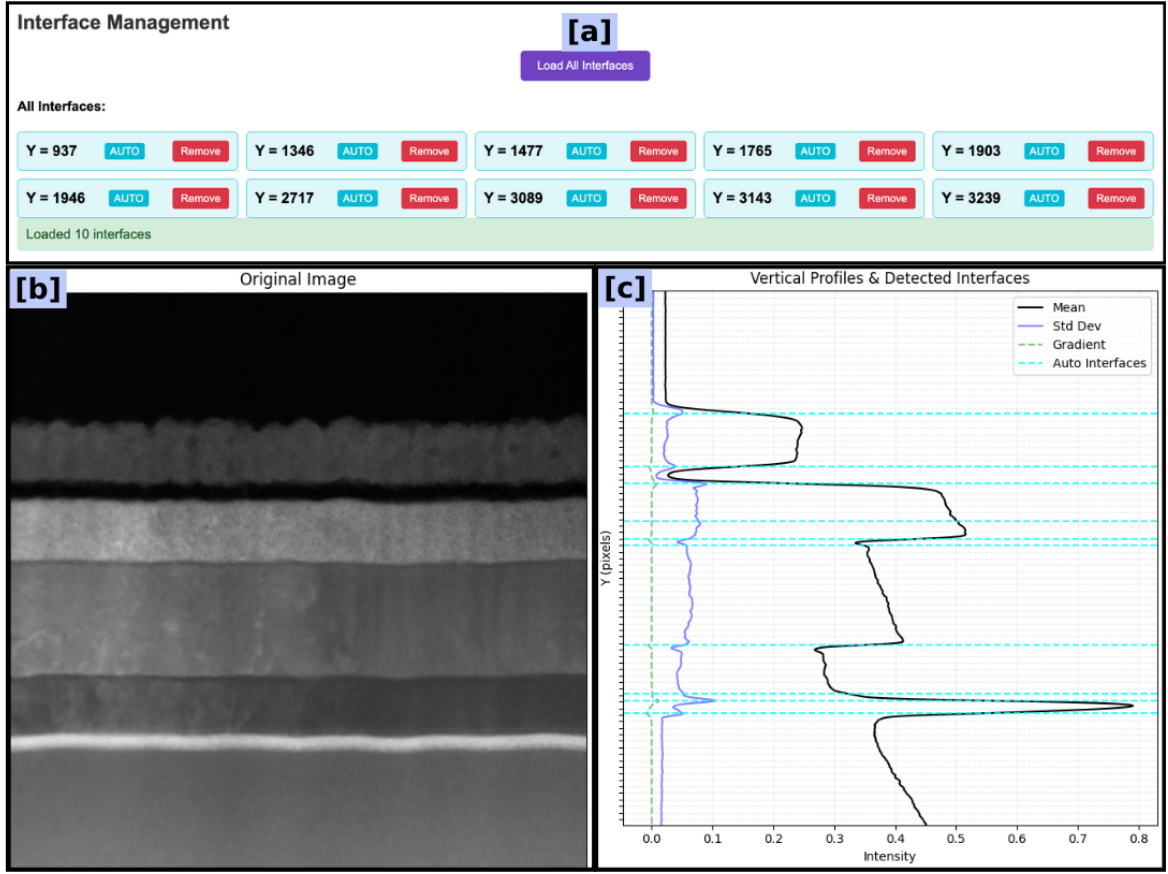

**Figure. 2:** Snippet of Web-based tool interface for real-time analysis of a multilayer TEM image. The display includes (a) the original image and (b) vertical intensity profiles with gradient analysis. A top panel lists all detected interfaces, where users can interact by removing existing interfaces or adding new ones. Cyan lines indicate automatically detected interfaces, when user interacts and adds an interface it is shown in red color for easy visual differentiation b/w manual and automatically found interfaces

### III A. User Interface and Functionality

- The tool provides a three-panel view (**Figure 2**): original image, vertical profiles, and interface management.
- Users can add/remove interfaces interactively (**Figure 2a**), with cyan marking automatic detection and red marking manual edits.

- Calibration metadata from .emd files is automatically parsed.

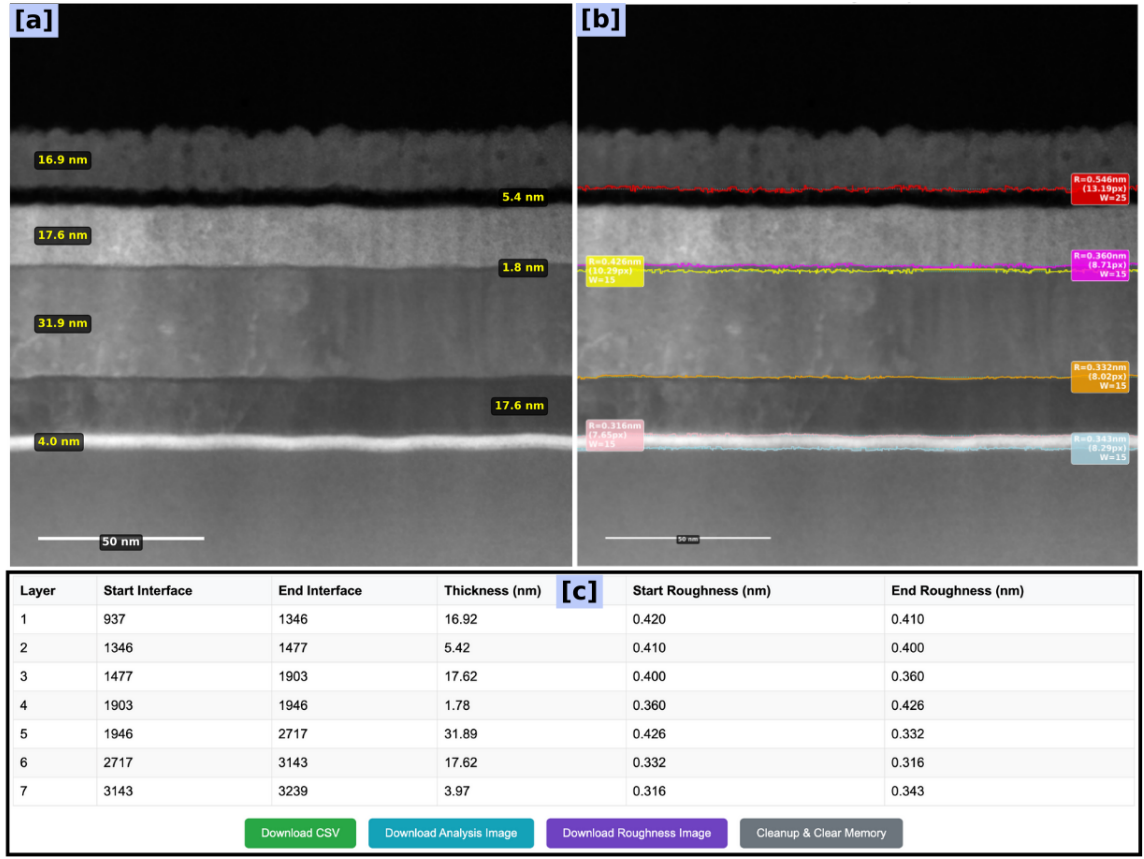

Figure. 3: Computed roughness and thickness measurement in the tool with export options. a) The thickness annotated label for the layer, b) shows calculated roughness for different layers. Panel c) shows the tabular data of (a) and (b) which can be exported as an excel file.

### III B. Measurement Accuracy and Validation

- Thickness estimation achieves nanometer-scale accuracy when calibration is available (**Figure 3a**).
- Interface roughness is quantified across multiple layers (**Figure 3b**), extending TEM analysis beyond traditional contrast inspection.
- Tabulated outputs (Figure 3c) provide a reproducible summary that can be exported for further analysis.
- Validation is restricted to visual comparison with expert inspection (**Figure 2b–c**).

- The method complements but does not replace AFM; it uniquely enables buried interface roughness analysis directly from TEM.

*Note the validation involved comparison with human annotated ground truth. Due to proprietary reasons those are not shared here.*

**III C. Comparison with Traditional Methods**

- Unlike AFM, which requires sequential surface access, this TEM approach allows simultaneous analysis of buried and surface interfaces (**Figures 2–3**).
- Automation significantly reduces operator bias and shortens analysis time.
- Standardized exports (Figure 3c) support reproducibility across different users and institutions.

**III D. Limitations and Future Developments**

- Current dependence on .emd format and 2D-only analysis limits applicability.
- Planned improvements: support for additional file formats, 3D tomographic integration, and ML-based interface recognition.
- Web-based architecture ensures seamless updates and facilitates deployment in SEM tools and potential manufacturing pipelines.
- Adding backend database support for scaling.

**IV. Conclusion**

We have developed a comprehensive AI-assisted web-based tool for automated analysis of multilayer structures from TEM images. The tool combines sophisticated image processing algorithms with intuitive user interfaces to provide accurate thickness and roughness measurements. Key contributions include:

- Automated interface detection with interactive manual correction capabilities
- Comprehensive roughness quantification using geometric tracking methods
- Web-based implementation ensuring universal accessibility
- Real-time visualization and comprehensive data export functionality
  This tool addresses critical needs in materials science research and industrial quality control, providing standardized, objective analysis of multilayer structures. The open architecture supports future enhancements and collaborative development within the materials science community.

The availability of automated TEM-based roughness analysis fills an important gap in materials characterization, complementing traditional AFM measurements and enabling comprehensive analysis of buried interfaces in complex multilayer systems.

## V. Acknowledgments

(UP, SVK) acknowledge support from Western Digital corporation.

## References

1. W. Xu et al., "AI-Powered Next-Generation Technology for Semiconductor Optical Metrology: A Review," Micromachines (Basel). **16**(8), 838 (2025) [doi:10.3390/mi16080838].

2. U. Celano et al., "Metrology for 2D materials: a perspective review from the international roadmap for devices and systems," Nanoscale Adv. **6**(9), 2260–2269 (2024) [doi:10.1039/D3NA01148H].

3. KLA, "Defect Inspection and Review," https://www.kla.com/products/chip-manufacturing/defect-inspection-review.

4. J. Severi et al., "Chemically amplified resist CDSEM metrology exploration for high NA EUV lithography," Journal of Micro/Nanopatterning, Materials, and Metrology **21**(02) (2022) [doi:10.1117/1.JMM.21.2.021207].

5. S. J. Haigh et al., "Cross sectional STEM imaging and analysis of multilayered two dimensional crystal heterostructure devices," Microscopy and Microanalysis **21**(S3), 107–108 (2015) [doi:10.1017/S1431927615001336].

6. R. F. Egerton and M. Malac, "EELS in the TEM," J. Electron Spectros. Relat. Phenomena **143**(2–3), 43–50 (2005) [doi:10.1016/j.elspec.2003.12.009].

7. J. Tröger et al., "Optimizing time-of-flight secondary ion mass spectrometry depth profiles of semiconductor heterostructures," J. Appl. Phys. **137**(2) (2025) [doi:10.1063/5.0232252].

8. Y. Zhou et al., "OrbiSIMS depth profiling of semiconductor materials—Useful yield and depth resolution," Journal of Vacuum Science & Technology A **42**(5) (2024) [doi:10.1116/6.0003821].

9. S.-B. Yoo et al., "Automated measurement and analysis of sidewall roughness using three-dimensional atomic force microscopy," Appl. Microsc. **52**(1), 1 (2022) [doi:10.1186/s42649-022-00070-5].

10. D. B. Williams and C. B. Carter, "The Transmission Electron Microscope," in Transmission Electron Microscopy, pp. 3–17, Springer US, Boston, MA (1996) [doi:10.1007/978-1-4757-2519-3_1].

11. M. Shen et al., "A deep learning based automatic defect analysis framework for In-situ TEM ion irradiations," Comput. Mater. Sci. **197**, 110560 (2021) [doi:10.1016/j.commatsci.2021.110560].

12. J. Madsen et al., "A Deep Learning Approach to Identify Local Structures in Atomic-Resolution Transmission Electron Microscopy Images," Adv. Theory Simul. **1**(8) (2018) [doi:10.1002/adts.201800037].

13. C. Gui et al., "Deep learning analysis on transmission electron microscope imaging of atomic defects in two-dimensional materials," iScience **26**(10), 107982 (2023) [doi:10.1016/j.isci.2023.107982].

14. I. Kojima, B. Li, and T. Fujimoto, "High resolution thickness and interface roughness characterization in multilayer thin films by grazing incidence X-ray reflectivity," Thin Solid Films **355–356**, 385–389 (1999) [doi:10.1016/S0040-6090(99)00544-1].

15. C. J. Tavares et al., "Study of roughness in Ti0.4Al0.6N/Mo multilayer structures," Nucl. Instrum. Methods Phys. Res. B **188**(1–4), 90–95 (2002) [doi:10.1016/S0168-583X(01)01026-6].

16. H. Jiang et al., "Determination of the evolution of layer thickness errors and interfacial imperfections in ultrathin sputtered Cr/C multilayers using high-resolution transmission electron microscopy," Opt. Express **19**(12), 11815 (2011) [doi:10.1364/OE.19.011815].

17. O. Filies et al., "Surface roughness of thin layers—a comparison of XRR and SFM measurements," Appl. Surf. Sci. **141**(3–4), 357–365 (1999) [doi:10.1016/S0169-4332(98)00524-8].

18. Y. Liu et al., "Integration of Scanning Probe Microscope with High-Performance Computing: fixed-policy and reward-driven workflows implementation" (2024).

19. K. Barakati et al., "Unsupervised Reward-Driven Image Segmentation in Automated Scanning Transmission Electron Microscopy Experiments" (2024).

20. I. Sokolov, "On machine learning analysis of atomic force microscopy images for image classification, sample surface recognition," Physical Chemistry Chemical Physics **26**(15), 11263–11270 (2024) [doi:10.1039/D3CP05673B].

21. S. Lu et al., "Semi-supervised machine learning workflow for analysis of nanowire morphologies from transmission electron microscopy images," Digital Discovery **1**(6), 816–833 (2022) [doi:10.1039/D2DD00066K].

22. M. Ziatdinov et al., "AtomAI framework for deep learning analysis of image and spectroscopy data in electron and scanning probe microscopy," Nat. Mach. Intell. **4**(12), 1101–1112 (2022) [doi:10.1038/s42256-022-00555-8].

23. M. Klinger, "More features, more tools, more *CrysTBox*," J. Appl. Crystallogr. **50**(4), 1226–1234 (2017) [doi:10.1107/S1600576717006793].

24. A. Kirillov et al., "Segment Anything" (2023).

25. A. Archit et al., "Segment Anything for Microscopy," Nat. Methods **22**(3), 579–591, Nature Publishing Group US, New York (2025).

26. W. Abebe et al., "SAM-I-Am: Semantic boosting for zero-shot atomic-scale electron micrograph segmentation," Comput. Mater. Sci. **246**, 113400, Elsevier B.V, United States (2025).

27. N. Kulesh et al., "Data-driven optimization of FePt heat-assisted magnetic recording media accelerated by deep learning TEM image segmentation," Acta Mater. **255**, 119039 (2023) [doi:10.1016/j.actamat.2023.119039].

28. K. Nichols et al., "Segment anything model for grain characterization in hard drive design," in Proceedings of the IEEE/CVF Conference on Computer Vision and Pattern Recognition, pp. 8120–8124 (2024).

29. U. Pratiush et al., "Building Workflows for Interactive Human in the Loop Automated Experiment (hAE) in STEM-EELS" (2024).

30. X. Wu et al., "A survey of human-in-the-loop for machine learning," Future Generation Computer Systems **135**, 364–381 (2022) [doi:10.1016/j.future.2022.05.014].

31. S. V Kalinin et al., "Human-in-the-Loop: The Future of Machine Learning in Automated Electron Microscopy," Micros. Today **32**(1), 35–41 (2024) [doi:10.1093/mictod/qaad096].

# Supplementary

## SI A. Image Preprocessing and Enhancement

The analysis pipeline begins with image normalization to ensure consistent processing across different imaging conditions:

$$I_{norm}(x,y) = \frac{I(x,y) - I_{\min}}{I_{\max} - I_{\min}} \quad (1)$$

where $I(x,y)$ represents the original intensity at pixel coordinates ($x,y$), and $I_{norm}(x,y)$ is the normalized intensity in the range [0,1].

Gaussian smoothing is subsequently applied to reduce noise while preserving interface information:

$$I_{\text{smooth}}(x,y) = I_{\text{norm}}(x,y) * G_\sigma(x,y) \quad (2)$$

where $G_\sigma(x,y)$ is a 2D Gaussian kernel with useradjustable standard deviation $\sigma$. The convolution operation (*) effectively reduces high-frequency noise while maintaining edge definition at interfaces.

## SI B. Interface Detection Algorithm

Interface detection is performed through analysis of vertical intensity profiles. For each horizontal position $x$, we compute:

$$P_{\text{mean}}(y) = \frac{1}{W} \sum_{x=1}^{W} I_{\text{smooth}}(x,y) \quad (3)$$

$$P_{\text{grad}}(y) = \frac{dP_{\text{mean}}(y)}{dy} \quad (4)$$

where $W$ is the image width and $P_{\text{grad}}(y)$ represents the vertical gradient profile that highlights interface locations.

Automated peak detection utilizes a percentile-based threshold approach:

$$\tau = \text{percentile}\left(|P_{\text{grad}}|, p\right) \tag{5}$$

where $p$ is the percentile threshold (typically 95%) and $\tau$ is the resulting height threshold for peak detection. This adaptive approach accommodates varying contrast conditions across different samples.

Peak detection is performed using the SciPy find_peaks algorithm with distance constraints:

$$\text{Peaks} = \text{find\_peaks}\left(|P_{grad}|, \text{height} = \tau, \text{distance} = d_{\min}\right) \tag{6}$$

where $d_{\min}$ is the minimum allowed distance between adjacent peaks, preventing false detection of noise-induced peaks.

**SI C. Interface Roughness Quantification**

Interface roughness is calculated using geometric tracking methods that analyze interface position variations across the image width. For each interface at nominal position $y_0$, we define an adaptive search window:

$$W_{\text{search}} = \min\left(W_{\text{base}}, \frac{1}{2}\min(|y_{\text{prev}} - y_0|, |y_{\text{next}} - y_0|)\right) \tag{7}$$

where $W_{\text{base}}$ is the base window size (15 pixels for most interfaces, 25 pixels for the second interface based on empirical optimization), and $y_{\text{prev}}$, $y_{\text{next}}$ are adjacent interface positions.

Within each vertical column $x$, the actual interface position $y_{\text{actual}}(x)$ is determined by finding the maximum absolute gradient:

$$y_{\text{actual}}(x) = y_0 + \arg\max_{i \in [-W_{\text{search}}, W_{\text{search}}]} |P_{\text{grad}}(y_0 + i, x)| \tag{8}$$

The interface deviation from the nominal position is calculated as:

$$\delta(x) = y_{\text{actual}}(x) - y_0 \tag{9}$$

Root mean square (RMS) roughness is computed as:

$$R_{RMS} = \sqrt{\frac{1}{W}\sum_{x=1}^{W} \delta(x)^2} \tag{10}$$

Additional roughness parameters include average roughness:

$$R_a = \frac{1}{W}\sum_{x=1}^{W} |\delta(x)| \tag{11}$$

and maximum peak-to-valley roughness:

$$R_{\max} = \max(\delta) - \min(\delta) \tag{12}$$

## SI D. Thickness Calculation

Layer thickness is calculated as the distance between consecutive interfaces:

$$t_i = (y_{i+1} - y_i) \cdot s_{\text{pixel}} \tag{13}$$

where $t_i$ is the thickness of layer $i$, $y_i$ and $y_{i+1}$ are the vertical positions of consecutive interfaces, and $s_{\text{pixel}}$ is the pixel size in nanometers extracted from EMD metadata.

Statistical analysis provides comprehensive characterization:

$$\bar{t} = \frac{1}{N} \sum_{i=1}^{N} t_i \tag{14}$$

$$\sigma_t = \sqrt{\frac{1}{N-1} \sum_{i=1}^{N} (t_i - \bar{t})^2} \tag{15}$$

where $N$ is the number of layers, $\bar{t}$ is the mean thickness, and $\sigma_t$ is the standard deviation.

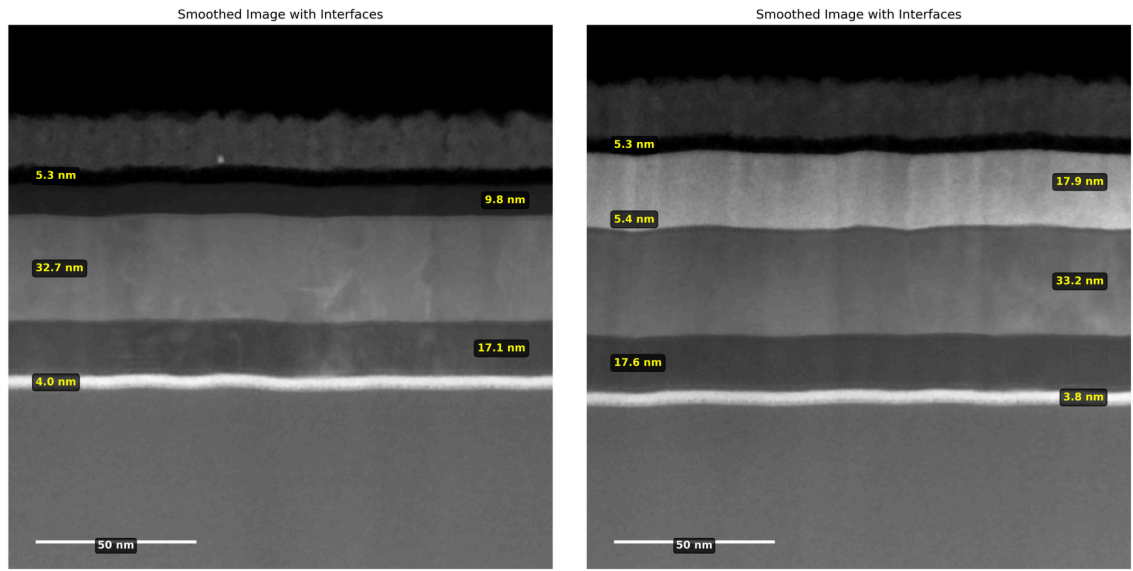

**Figure S1a.** Example of thickness measurements on more multilayer model structures.

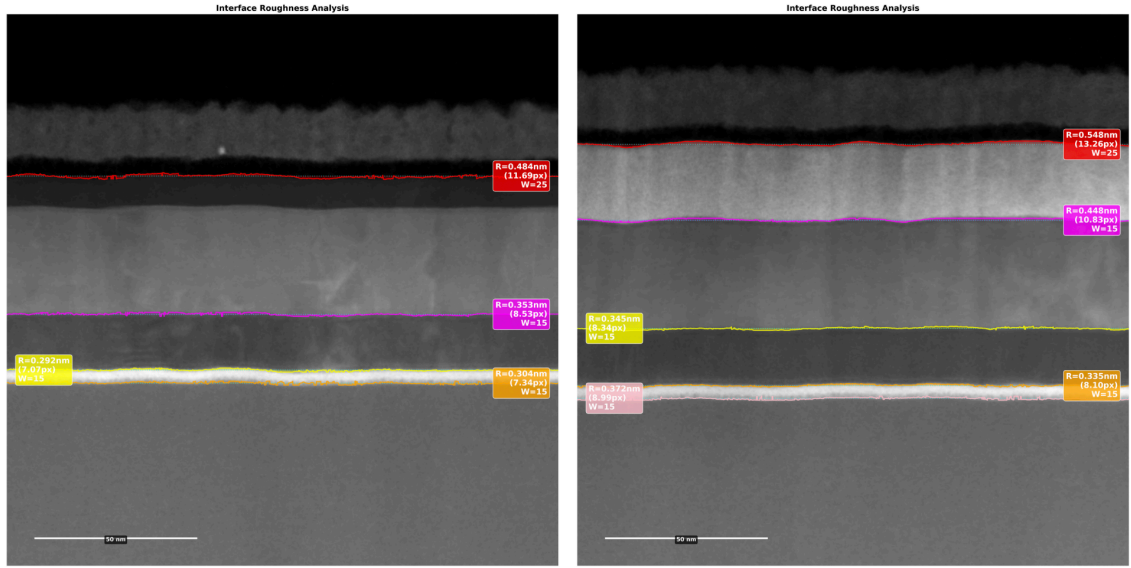

**Figure S1b.** Example of roughness measurements on more multilayer model structures.